\documentclass[aps,pre,twocolumn,showpacs,preprintnumbers,amsmath,amssymb,superscriptaddress]{revtex4}
\usepackage{graphicx}
\usepackage{bm}
\usepackage{psfrag}
\usepackage{subfigure}
\usepackage{color}

\newcommand{\be}{\begin{equation}}
\newcommand{\ee}{\end{equation}}
\newcommand{\ba}{\begin{eqnarray}}
\newcommand{\ea}{\end{eqnarray}}

\newcommand{\mec}[1] {\textcolor{black}{#1}}
\newcommand{\ma}[1] {\textcolor{black}{#1}}

\begin{document} 
\author{Matthieu Wyart}

\affiliation{Institute of Physics, EPFL, CH-1015 Lausanne, Switzerland} \author{Michael E. Cates}
\affiliation{DAMTP, Centre for Mathematical Sciences, Wilberforce Road, Cambridge, CB3 0WA, United Kingdom}

\date{\today}

\title{Does a growing static length scale control the  glass transition?}

\begin{abstract}

Several theories of the glass transition propose that the structural relaxation time $\tau_\alpha$ is controlled by a growing static length scale $\xi$ that is determined by the free energy landscape but not by the local dynamical rules governing its exploration. We argue, based on recent simulations using particle-radius-swap dynamics, that only a modest factor in the increase in $\tau_\alpha$ on approach to the glass transition may stem from the growth of a static length, with a vastly larger contribution attributable instead to a slowdown of local dynamics. This reinforces arguments  that we base on the observed strong coupling of particle diffusion and density fluctuations in real glasses. 
\end{abstract}


\maketitle

When a liquid is cooled sufficiently rapidly to avoid crystallization, its viscosity  $\eta$ increases. In general, 
$\eta\simeq G \tau_\alpha$, where $G$ is the \mec{(plateau)} shear modulus  and $\tau_\alpha$ characterizes the relaxation time of density fluctuations. As the temperature $T$ is lowered, $G$ evolves mildly but $\tau_\alpha$ increases by about $15$ order of magnitude \cite{Ediger96} until, at the transition temperature $T_G$, it becomes too large to measure experimentally. The liquid then becomes a glass, and falls out of equilibrium. Near the glass transition, the diffusion of a tagged particle also becomes very slow. The characteristic time $\tau_D$ over which a particle diffuses its own radius increases, albeit not as much as $\tau_\alpha$. The decoupling of these two quantities (referred to as the Stokes-Einstein breakdown) is significant, but comparatively mild: the ratio ${\cal S} = \tau_\alpha/\tau_D$ is increased by only a few orders of magnitude at $T_G$ \cite{Ediger00,Swallen09}. 

The dependence of $\tau_\alpha$ on $T$ is used to classify glassy liquids \cite{Angell91}. Writing $\tau_\alpha=\tau_0 \exp(E/k_BT)$ where $\tau_0$ is a vibrational time scale (in the picosecond range) and  $E$ some activation energy, one finds that $E$ is constant in some liquids, called strong, but  increases up to a factor $5$ under cooling in other liquids, called fragile.  Fragility is best shown in the `Angell plot' of $\log(\tau_\alpha)$ {\it vs.} $T_G/T$ \cite{Angell91}, which is linear for strong liquids but highly curved for fragile ones. Quantitatively, fragility is defined as $m=d \log(\tau_\alpha)/d \log(T_G/T)$ evaluated at $T_G$ itself. Strong liquids have $m\approx 25$; very fragile ones have $m\approx 120$. 

There are competing explanations of the increase in the activation energy $E$  in fragile liquids, which occurs with no obvious change of static structure \cite{Berthier11b}. Several theories, including the Adam-Gibbs scenario \cite{Adam65}, Random First Order Theory (RFOT) \cite{Lubchenko07,Bouchaud04,Kirkpatrick15}, and those involving locally favored structures \cite{Kivelson95,Tanaka12}, posit that the increase in $E$ stems from the growth of a purely static length scale $\xi$, characterizing some `hidden order' in the many-body free-energy landscape that is not captured by traditional probes of static structure such as pair correlations. 

In particular, in modern interpretations of RFOT \cite{Lubchenko07,Bouchaud04}, $\xi$ is a `point-to-set' correlation length set by the minimum scale on which alternative packings are available to a patch of fluid whose environment is held frozen. Shorter scale motions do cross local barriers, but cannot discover a new pattern for the density which keeps returning to its initial state. In this view, regardless of how rapidly these local moves can permute the particles within the patch, $\tau_\alpha$ is controlled by the fact that the system can relax \ma{fully} its density fluctuations only via collective rearrangements on the scale $\xi$ that `break' the hidden order. 
The resulting collective activation energy is $ E_{\rm coll} \simeq c_0(T) \xi(T)^\psi$, where $c_0(T)\simeq c_0(T_G)$ is non-singular, and $\psi$ is some exponent. \ma {It can be expressed in terms of thermodynamic quantities alone and is thus independent of  the details of the dynamics \footnote[1]{\ma{This statement holds true if the order parameter $Q$ on a correlation volume $\xi^d$ evolves smoothly at each numerical step, as it ensures that configurations with  $Q=Q^*$ for which the free energy barrier is maximum are visited. This must be true  for any Monte-Carlo where moves involves $n_0$ particles with  $n_0<<\xi^d$. \label{footnote1}}} (see more on that below).}

Although there is clear empirical evidence that a growing static length scale $\xi(T)$ exists (see e.g. \cite{Albert16} and references therein), its role  in the dynamics is debated. Alternative theories instead propose that the increase of $\tau_\alpha$ stems from growing barriers to the \ma{elementary rearrangements required to explore the landscape \cite{Dyre06,Torchinsky09,Mirigian14, Garrahan11,Lemaitre14}}. Such barriers effectively add a term to the activation energy $\ma{E_{\rm el}} = c_1(T)$. \ma{ Unlike $E_{\rm coll}$, these kinetic barriers  can  depend greatly on the details of the dynamical rules governing the system.} While the definition of `\ma{elementary rearrangements}' can be \ma{multi-particle}  (see more on that below), $\ma{E_{\rm el}}$ describes local physics and cannot diverge. However it might grow strongly enough near $T_G$ to control the glass transition via $\tau_\alpha\propto\ma{\tau_{\rm el}} \simeq \tau_0\exp[\ma{E_{\rm el}}/k_BT]$. For example in elastic models \cite{Dyre06}, rearrangements require a certain strain, giving  $E(T)\sim G_\infty(T)$ where $G_\infty$ is the \mec{high-frequency (vibrational) shear modulus}.  Empirically these quantities are indeed strongly correlated \cite{Dyre06,Torchinsky09}, as seen by plotting $\log(\tau_\alpha)$ {\it vs.} $(T_G G_\infty(T))/(T G_\infty(T_G))$, which now appears almost perfectly linear even for fragile liquids \cite{Torchinsky09,Hecksher15}.

Combining the two activation terms gives $\ln(\tau_\alpha/\tau_0) \simeq a_0\xi^\psi + a_1$, with $a_0 = c_0/k_BT$ and $a_1=c_1/k_BT$. This estimate coincides in form with an upper bound on the relaxation time \cite{Montanari06a}; this bound shows that a {\em diverging} $\xi$ is rigorously required for an ideal glass transition ($\tau_\alpha/\tau_0\to \infty$). It is then tempting to conclude that {\em growth} of $\xi$ controls the real glass transition ($\tau_\alpha/\tau_0\to 10^{15}$). 

\begin{figure}[htbp]
\centering
\includegraphics[width=1\columnwidth]{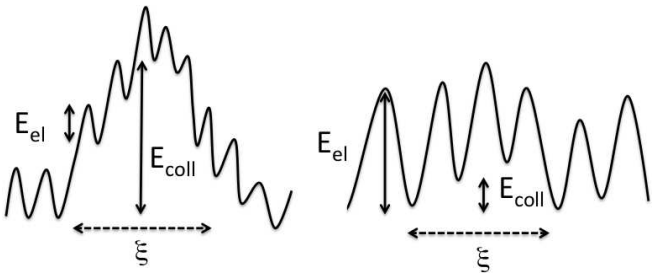}
\caption{\mec{Left panel: schematic of the free energy landscape $F(\{{\bf r}\})$} for models where the dominant kinetic barriers $E_{\rm coll}$ are collective, and appear on a length scale $\xi$. Local barriers $\ma{E_{\rm el}}$ are much smaller than $E_{\rm coll}$. In this \mec{scenario introducing swap moves in the dynamical rules should have little effect on \mec{structural relaxation}, since it leaves $E_{\rm coll}$ unchanged; also} there should be a very large decoupling between particle diffusion (that can occur by permuting particles, which requires to overcome local barriers only) and relaxation of density fluctuations (for which collective barriers must be overcome). Both predictions appear to be contradicted by observations, \mec{suggesting that  local barriers dominate the slowdown (right panel).} \ma{(In this case the definition of $E_{\rm coll}$ and $\xi$ may become somewhat arbitrary.)} }  \label{f1}
\end{figure}

In this work we present arguments for the contrary view, proposing instead that most of the increase in $\tau_\alpha$ stems from an increase upon cooling of $\ma{E_{\rm el}}$. Our main argument is based on  recent numerical observations \cite{Berthier16b,Gutierrez15,Ninarello17,Berthier17} in which a judicious choice of dynamics is shown to equilibrate systems far deeper inside the glass \mec{($T\ll T_G$)} than previously achievable. Indeed this dynamics almost abolishes the glass transition. We shall infer from this outcome that $\ma{E_{\rm el}}$ rather than $E_{\rm coll}$ is dominant in controlling $\tau_\alpha$. This finding applies, strictly speaking, only to the polydisperse systems studied numerically in \cite{Berthier16b,Gutierrez15,Ninarello17,Berthier17}. %; yet these appear to be fully representative of real glasses when the standard dynamics is restored. 
We then give a more general and complementary argument, to the effect that the Stokes-Einstein factor \mec{should obey ${\cal S} \ge \exp[E_{\rm coll}/k_BT] = \tau_\alpha/\ma{\tau_{\rm el}}$}. It follows that the growth of $\xi$ contributes only a few decades at most to the 15 decades increase in $\tau_\alpha$ on the approach to $T_G$. \mec{Our findings are depicted schematically in Fig.\ref{f1}.}

{\em Swap algorithms.}  Recent Monte-Carlo schemes have managed to equilibrate liquids at temperatures where  traditional algorithms would need at least a $10^{10}$ speed-up to allow equilibration \cite{Grigera01b,Fernandez06, Berthier16b,Gutierrez15,Ninarello17,Berthier17}. These swap algorithms consider polydisperse particles. In addition to the usual translational moves, they allow for moves in which two particles of different radius exchange places. 
This can also be viewed as a radius-swap at fixed particle positions (in which case there is no direct contribution to tagged-particle diffusivity). One then has particles whose sizes are an extra set of dynamical variables, with update rules that preserve detailed balance. Accordingly the free energy landscape $F(\{{\bf r}\})$ as a function of the particle positions $\{{\bf r}\}$ is, after integration over sizes,  well defined and independent of the chosen dynamical rules. % \footnote[2]{A similar averaging over polydispersity is implicit in measurements of dynamic structure factors in colloidal glasses \cite{Urban98}.}.  
The swap rules can involve either well-separated particles as reported in \cite{Ninarello17,Berthier17}, or be local, for instance restricted to nearest neighbour swaps  \cite{Berthier17p}. For the dynamical rules of \cite{Ninarello17,Berthier17}, the time scale $\tau_\alpha^+$ on which density fluctuations relax with swaps \footnote[5]{ \mec{Note that in  \cite{Berthier17}, $\tau_D$ and $\tau_\alpha$ are both treated as one-particle properties. Here we use $\tau_\alpha$ to describe collective density relaxations.}}  can be measured and compared to the value $\tau_\alpha^-$ without swaps allowed. The latter case behaves like molecular dynamics simulations, and display the standard glass transition. Remarkably though, with the swap dynamics switched on the glass transition is essentially gone: $\tau_\alpha^+(T_G)/\tau_0$ is only around $10^2-10^3$. This contrasts with $\tau_\alpha^-/\tau_0\simeq 10^{15}$ for the standard dynamics at $T_G$. % as normally defined\cite{Ediger96}.  

{\em Implications of these observations.} 
%We  assume that local and nonlocal swap dynamics are equivalent for our purposes, explaining why this should be so here \footnote[4]{Local and long-range swaps  should be equivalent for our purposes at least asymptotically, because if indeed the glass transition is controlled by a growing static length, local radius-swaps are already enough to fully anneal the particle size distribution within a patch of fluid of size $\xi$. Whereas for infinite-range swaps the time scale for this equilibration is of order one swap per particle, for local swaps it is diffusive, of order $\xi^2$. But \mec{when $E_{\rm coll}\gg \ma{E_{\rm el}}$} this is still negligible compared to $\tau_\alpha\sim\exp[a_1\xi^\psi]$. Therefore if relaxation in a radius-swap system is controlled by a collective barrier $E_{\rm coll}$, the landscape that determines this barrier must be the size-annealed one $F(\{{\bf r}\})$, even if the swaps are local. This argument holds asymptotically for large $\xi$, whereas in practice $\xi$ remains modest near $T_G$ \cite{Albert16}. }.
%
\ma{ Local and long-range swap dynamics do appear to give very similar results for $\tau_\alpha$ \cite{Berthier17}, as we justify in S.I.} This suggest that either type of swap dynamics \mec{can} mollify local barriers, leading to much faster relaxation than the conventional dynamics of real glasses. It follows that the latter is dominated by $\ma{E_{\rm el}}$.
\ma{Indeed}, switching from standard to swap dynamics simply enhances the local move set: swapping is a specific local move involving two particles only \footnote[3]{A different instance of local dynamical rules profoundly affecting the glass transition is discussed in \cite{Jacquin15}.}. The underlying free-energy landscape $F(\{{\bf r}\})$ is unaffected. As such, so long as one first equilibrates the system, any static ({\em i.e.}, thermodynamic) length $\xi$ must be identical with both dynamics. Indeed, a major selling-point of the swap algorithms is that they {\em do allow full equilibration} over an unprecedentedly wide temperature range.
We further consider \ma{(see  proof below)} that for large enough $\tau_\alpha$ the collective barriers $E_{\rm coll}$ must be identical for swap and non-swap dynamics. We can now compare the dynamics with ($+$) and without ($-$) swaps. 
Since $\tau_\alpha^\mp = \ma{\tau_{\rm el}}^\mp\exp[E_{\rm coll}(\xi)/k_BT]$, if glass physics is dominated by the growing length scale $\xi$, then $\tau_\alpha^{-}/\tau_\alpha^{+}$ should depend only weakly on $T$: indeed the only such dependence is through the factor $\ma{\tau_{\rm el}}^{-}/\ma{\tau_{\rm el}}^{+}$.  But in the simulations, $\tau_\alpha^{-}/\tau_\alpha^{+}$ increases without apparent limit as $T$ is lowered: this increase can be tracked for about 4 decades before it becomes too large to measure \cite{Ninarello17,Berthier17}. In contrast, the growth of $\tau_\alpha^+$, which \mec{bounds above} the slowing down caused by the growth of $\xi$, only shows a 2 or 3 decade increase at $T_G$. However, it continues to grow by several more decades as $T$ is further decreased. 

We offer the following interpretation for these results. The relaxation time $\tau_\alpha^+$ describing the with-swap dynamics, for which local barriers are mollified, may ultimately be controlled by the growth of $\xi$, \ma{although other effects are plausible}   %if so, its continued growth for $T \ll T_G$ is  laid open to investigation by these simulations
~\footnote[33]{\ma{One may expect a transition for swap dynamics at some temperature $T_{MCT}^{+}$ where local minima appear in $F(\{{\bf r}\})$ with $T_{MCT}^{-}> T_{MCT}^{+} $. Indeed for the swap dynamics stability is more demanding since radii can smoothly evolve (if poly-dispersity is continuous).}}.
%as well, where the particle radii are considered as dynamic variables  in addition to the particles positions. The restoring force acting on the radius $R$ would then be given by the $R$-dependent chemical potential $\mu(R)$ reflecting the system polydispersity, \mec{which would make our upper bound even tighter.}}. 
However, with the swap-free dynamics relevant to real glasses, the resulting contribution to $\tau_\alpha^-/\tau_0$ is smaller than 3 decades at the glass transition. Thus the lion's share of the growth in $\tau_\alpha$ stems not from a growing static length but from growing local barriers. Only these barriers are affected by introducing local swaps, so that the resulting collapse of the glass transition points directly to their dynamical importance.

 {\em Stokes-Einstein violation.} The arguments above are restricted to polydisperse systems, as simulated by radius-swap
 algorithms. However for conventional swap-free dynamics the relaxation in these systems is no different from classical numerical models of structural glasses \cite{Ninarello17,Berthier17}. This suggests that polydispersity is immaterial to the issue of whether a growing static length controls the glass transition. To support this view, we now present a polydispersity-free argument that again points toward control predominantly by local barriers. 
 
 For concreteness we suppose RFOT to correctly identify the static length scale $\xi$. Recall that in this approach, each configuration is a mosaic  of states, whose characteristic size $\xi$ results from a competition between configurational entropy and ``surface tension'' between states. The activation barrier to nucleate a new \mec{density configuration} on a length scale $r$ then varies as $c_0 r^\psi$: the dynamics is {\it fast} on short length scales, but slow on long length scales. However for $r<\xi$, a single state is thermodynamically favored. Local rearrangements are possible, but these will not fully relax density fluctuations.   Processes that can relax density fluctuations  occur on a scale $\xi$, the smallest on which alternative density patterns can appear \cite{Bouchaud04}. 

\mec{We believe that this picture contradicts the observation that in real glasses the Stokes Einstein (SE) factor ${\cal S} = \tau_\alpha/\tau_D$ at $T_G$ is of order $10^3$ \cite{Swallen11,Swallen09}, rather than a much larger value. In practice, much of the violation is thought to arise from dynamic heterogeneity (DH), in which diffusive and structural relaxation times are dominated by the most liquid and most solid regions respectively \cite{Tarjus95}. However, DH can only increase ${\cal S}$: any other mechanism found to contribute to the SE violation therefore gives a lower bound on it. We may therefore write ${\cal S}={\cal S}_1{\cal S}_2\ge{\cal S}_1$, where ${\cal S}_2$ accounts for DH, and ${\cal S}_1$ accounts for the possibility that particles can exchange positions while leaving density fluctuations unchanged \cite{Charbonneau14h}. 
We now argue that ${\cal S}_1\sim \exp[a_1\xi^\psi]$, by noting first that, even for $r<\xi$, there is always an exponentially large number of states available for conventional dynamics, corresponding to permutations of the particles within a fixed density pattern. Any such state can be reached by a series of local permutations, which can each be generated by  local rearrangements only. Without polydispersity there {\em cannot} be any thermodynamic reason for a particular permutation to be preferred, so a tagged particle {\em can} move diffusively with $\tau_D\simeq \tau_0\exp[\ma{E_{\rm el}}/k_BT] = \tau_\alpha/{\cal S}_1$.}
 
\mec{We then have ${\cal S}_1 = \exp[a_1\xi^\psi]<{\cal S}\simeq 10^3$, implying that the growth of $\xi$ cannot account for more than} about a 3 decade increase in $\tau_\alpha$ at $T_G$. Were the glass transition controlled purely by growth of $\xi$, a much stronger decoupling of $\tau_D$ and $\tau_\alpha$ could be expected. This is incompatible with the fact that the fragility defined from $\tau_D$ is almost as large as that defined  from $\tau_\alpha$ (only 13-25\% \footnote[34]{This change of fragility can be deduced from the power law fit reported between $\tau_D$ and $\tau_\alpha$ in \cite{Swallen11,Swallen09}.} smaller for the liquids studied in \cite{Swallen11,Swallen09}). In practice therefore, diffusive and density relaxations are much more strongly coupled than RFOT seems to imply, and this points to a major role for $\ma{E_{\rm el}}$ at the glass transition.
 The same arguments apply to any theory for which fast local moves are insufficient to break a hidden order on a growing scale $\xi$. 
 
\mec{In the specific case of RFOT our argument can be restated as follows: for the nucleation picture to hold (whereby a new density pattern appears by an activated barrier crossing at scale $\xi$) local equilibrium in the landscape $F(\{{\bf r}\})$ should be reached, on all length scales shorter than $\xi$, at time scales much smaller than $\tau_\alpha$. But this local exploration of phase space permutes particles, leading to diffusion on the same fast timescale. Thus in any temperature regime where RFOT is dominant in driving the slowdown in $\tau_\alpha$, a severe Stokes Einstein breakdown should occur. This is not seen in practice.}

\mec{\emph{Asymptotic equivalence of landscapes.} We now return to the case of polydisperse particles and explain why, as we assumed above, the free energy landscape $F(\{{\bf r}\})$ is the same for dynamics with and without swaps, for large enough $\tau_\alpha$. This implies in turn that $E_{\rm coll}$ is the same (including its prefactor $c_1$) for the two types of dynamics, precisely in the regime in which a growing static length could come to dominate the dynamics. The argument is almost the same as the one just given for monodisperse particles: for large $\xi$, local rearrangements are rapid in comparison to $\tau_\alpha$, sampling with Boltzmann weight configurations where groups of particles are permuted locally. Thus traditional (non-swap) dynamics already performs local swaps on  time scales much shorter than $\tau_\alpha$ in this limit. The collective free energy barriers, when large, must thus be described by the same $F(\{{\bf r}\})$ with and without swaps, which is what we assumed above.}

%MW: Not sure of the relevance of this work for reasons I explained in my emails. 
%Finally, we note that not only the swap algorithms but also analytic calculations point to the possibility of almost abolishing the glass transition through a judicious change in the choice of local dynamical rules without altering the Hamiltonian [Wijland PRL]. 

%[add. : discussion on t diff/ tau alpha. again 3 decades for cooperativity]

{\em Discussion}. We are not suggesting that a growing length scale $\xi$ cannot contribute at all to the slowdown of dynamics in real glasses. But to play more than a supporting role, the resulting order must affect elementary rearrangements themselves, via $\ma{E_{\rm el}}$. For example, in the spirit of elastic models, locally favored structures that grow under cooling \cite{Kivelson95,Tanaka12} could increase $G_\infty$, which slows down local moves requiring finite strain. Similarly, the observation that radius-swap moves dramatically speed up the dynamics of glasses does not contradict dynamically-facilitated models \cite{Garrahan02,Toninelli07}, since in these models the choice of local dynamics is important. However they do need to \ma{be consistent with the fact that} particle diffusion and density relaxation dynamics are strongly coupled.

In our view, the proof that a static length scale must diverge if $\tau_\alpha$ does \cite{Montanari06a} is of limited practical relevance for real liquids near their glass transition, since the proper signal for this divergence is $\tau_\alpha/\tau_D$ which increases relatively mildly within the experimental range. (Moreover, $\xi$ need only diverge as $(\ln(\tau_\alpha/\tau_D))^{1/\psi}$.) The same conclusion is reached if, instead of considering the dynamics in terms of the rescaled temperature $T/T_G$, one normalizes it by a natural energy scale $G_\infty(T)/G_\infty(T_G)$ (which cannot be singular for smooth interactions). As stated earlier, the dynamics shows no sign of divergence if plotted in terms of $(T_G G_\infty(T))/(T G_\infty(T_G))$ \cite{Torchinsky09,Hecksher15}. 

On the other hand, with local barriers mollified by swap dynamics as in \cite{Ninarello17,Berthier17}, the growth of $\xi$ could become paramount, and it is possible that RFOT is the correct theory of such dynamics. %. If so, a fully general theory of glass transitions would need to combine this with a 
For realistic dynamics however theories of local barriers appear to be needed to describe the glass transition. 
One such theory is provided by elastic models \cite{Dyre06,Torchinsky09}, for which $\ma{E_{\rm el}}$ is proportional to $G_\infty$. However the spatial description of local moves remains very crude in these models. 
Importantly, they can involve finite collections of particles, for purely dynamical reasons. For example, it may require less strain energy to permute three particles forming a triangle than two neighboring particles. 

\ma{Understanding the geometry of such elementary rearrangements is a challenge.} Recently there has been a convergence of different approaches to compute the properties of soft collective vibrational modes in hard sphere liquids: a real space description \cite{Brito06,Brito09}, mean-field approximation \cite{DeGiuli14b,DeGiuli14} and  exact calculations in infinite dimensions \cite{Franz15,Altieri16,Maimbourg16}. (The discussion below may apply to other liquids by replacing the packing fraction $\phi$ by $T$.) It is found that  some elastic modes become stable  only above an onset packing fraction $\phi_0$  \cite{Brito09,Charbonneau14,Maimbourg16}.  For $\phi<\phi_0$, one expects flow to be fast as the system can relax without jumping over free energy barriers along unstable directions of phase space. By contrast, flow must be activated for $\phi>\phi_0$. This phenomenon is captured in part by the mode-coupling theory of liquids and by mean-field theory  \cite{Rabochiy13, Lubchenko07,Grigera02,Franz07,Andreanov09}. One expects  a growing {\it dynamical} length scale at such intermediate $\phi$ because the last modes to remain unstable as $\phi$ increases are increasingly collective.  For larger $\phi$, the simplest guess is that activated {\ma elementary rearrangements} occurs along similar modes to those whose stability is exchanged  at $\phi_0$. This is consistent with the observation that the dynamical length scale almost saturates (at the scale of collective motions of a few tens of particles) once $\tau_\alpha$ has increased by a few orders of magnitude \cite{Dalle-Ferrier07}.  

This line of thought is  consistent with the observations of swap algorithms. Indeed if \ma{elementary rearrangements} become more collective because local moves are too costly, then dynamical correlations \ma{can} disappear if \ma{effective} local moves such as radius-swaps are allowed for. Alongside much faster relaxation, dynamical correlations near the mode coupling temperature should then be eliminated. This is indeed observed: dynamical correlations become very small when swaps are used  \cite{Ninarello17}. Furthermore, dynamical correlations near $T_G$ also become very small. This observation undermines the RFOT result that dynamical heterogeneities near $T_G$  in liquids reflect $\xi$ \cite{Berthier11b}, since $\xi$ is unaffected by the choice of dynamics. \ma{ Instead it supports the view that for non-swap dynamics, elementary rearrangements are spatially more extended than $\xi$, further indicating that $T_G$ lies outside the temperature range where RFOT might apply.} 

Another point to clarify concerning elastic models  is why $G_\infty$ increases under cooling. From that perspective,  the exchange of stability of some elastic modes at $\phi_0$  can be shown to lead to an increase in shear modulus under cooling \cite{Lubchenko07,DeGiuli14,Nakayama15}, and is thus consistent with mean-field approaches. Other factors susceptible to stiffen the material  may be very system-specific, including the previously mentioned growth of locally-favored structures \cite{Kivelson95,Tanaka12}. %, an effect that would be very interesting to study.

A promising avenue that appears consistent with the swap-algorithm observations would thus be to combine elastic models with a more detailed description of collective dynamical modes beyond their linear regime. This could also account for other key facts of the dynamics in liquids, in particular the presence of growing {\em dynamical} length scales, the correlations between activation energy and high frequency shear modulus, and the correlations between entropy and dynamics \cite{Wyart10}.

{\em Conclusion}. The recent observation that swap algorithms can essentially eliminate the glass transition without changing the free energy landscape casts doubt on theories in which a growing static length scale, determined solely by this landscape and setting a minimum scale for finding new density patterns, controls the slowdown of relaxations in glasses. Instead, these observations cap the contribution from this source at a few decades of growth in the structural relaxation time, with a much larger factor arising instead from the growing barriers to local rearrangement. This concurs with a similar cap derived from the Stokes Einstein violation for real glasses which is likewise only a few orders of magnitude, requiring tagged particle diffusion (asymptotically unaffected by the growth of $\xi$) and structural relaxation to remain strongly coupled near the glass point. Among theories of the local barrier physics, a combination of elastic models with an improved description of collective dynamic modes may offer a promising route forward.

%{\em Acknowledgments}.

\begin{acknowledgements}
We thank G. Biroli, J.-P. Bouchaud, L. Berthier, J. Dyre, M. Ediger, J. Kurchan, H. Tanaka, D. Thirumalai and P. Wolynes for discussions.  MW thanks the Swiss National Science Foundation for support under Grant No. 200021-165509 and the Simons Foundation Grant ($\#$454953 Matthieu Wyart). MEC is funded by the Royal Society.
\end{acknowledgements}

\bibliography{../bib/Wyartbibnew}

\end{document}